# Trilayer embedded graphene flakes under efficient tensile and compressive loading

G. Tsoukleri1, J. Parthenios1, C. Galiotis1,2 and K. Papagelis1,3

*1Foundation of Research and Technology Hellas, Institute of Chemical Engineering and High Temperature Processes, P.O. Box 1414, GR-26504 Patras (Greece)*
*2Department of Chemical Engineering, University of Patras, GR-26504Patras (Greece)*
*3Department of Materials Science, University of Patras, GR-26504 Patras (Greece)*

**Abstract**
The mechanical response of embedded ABA trilayer graphene flakes loaded in tension and compression on polymer beams is monitored by simultaneous Raman measurements through the strain sensitivity of the G or 2D peaks. A characteristic peculiarity of the investigated flake is that it contains a trilayer and bilayer part. The Bernal stacked bilayer was used as a strain sensor aiming to assess the efficiency of the load transfer from the polymer matrix through shear to the individual graphene layers. For the trilayer graphene in tension, both peaks are redshifted and splitting of the G peak is reported for the first time. In compression, the studied sample was an almost isolated trilayer, in which both peaks are blue-shifted up to a critical compressive strain. This critical strain is found to be one fourth of the value found in the case of single layer graphene despite the higher bending rigidity that trilayer exhibits over the much thinner monolayer.

**1. Introduction**

It has been recently proved that the way, that single graphene lattices (1LG) are stacked together to form bi- (2LG) and trilayer (3LG) graphene crystals, leads to the formation of new materials with unique electronic properties which are considerably different from those of the monolayer graphene [1-3].
In contrast to monolayer graphene which is a zero-gap semimetal [4, 5], the AB stacked (Bernal) bilayer graphene is a semiconductor with a tunable gap [6]. By adding a third layer, the ABA-stacked (Bernal) trilayer graphene retains its metallic character under an applied electric field [1], while the ABC stacked trilayer graphene behaves in a similar manner to bilayer graphene, with a tunable semiconducting gap [1, 2]. However, as yet it has not been fully examined the effect of graphene stacking upon the mechanical properties in tension and compression. For instance, to our knowledge, only a few works have attempted to address the mechanical response of trilayer graphene either theoretically [7, 8] or experimentally [9]. Shen et al [7] using MD simulations investigated the interlayer shear effect in multilayer graphene subjected to bending. Recently Liu et al [8] have modelled the response of bi- and trilayer graphene under bending by taking into account the effects of the in-planeextension or contraction as well as the interlayer shear effect.

Raman spectroscopy is a fast, sensitive and non-destructive technique and has been already used successfully to determine the stacking order of trilayer graphene samples. For specific excitation energy, ABA and ABC stacked graphenes have been distinguished by their differences mainly regarding the full-width-at-half-maximum of the corresponding 2D Raman peaks [10, 11].

In the work of Gong et al. [9] the mechanical response of a trilayer graphene under axial tensile loading was examined through of the strain dependence of the 2D Raman peak. The maximum applied strain was 0.40%, and no evidence of G-mode splitting was observed which, for both 1LG and 2LG, is the key phenomenon to ensure that uniaxial strain conditions have been imposed to the flake. Finally, in a subsequent publication by Gong et al [12] a clear evidence of Bernal stacking sequence distortion in few layer graphene upon axial tensile strain has been presented. The alterations in the 2D Raman profile caused by moderate axial strains of up to 0.4% were used to identify the loss of Bernal stacking in few layer graphene structures.

In general, the application of tensile stress induces phonon softening (red shift) whereas compressive stress causes phonon hardening (blue shift). Under uniaxial stress/ strain conditions, the double degeneracy of the E2g phonon is lifted and the G peak splits in two components polarized parallel (G− mode) and perpendicular (G+mode) to the strain axis [13, 14]. In very recent past, a number of authors have combined Raman Spectroscopy and mechanical measurements in order to examine the behaviour of monolayer graphene flakes (1LG) under tension [9, 13-21] and compression [14, 15]. The results have shown that there is a small variation of the slopes of G−, G+ ( ranging between ∼30 – 34 and ∼10 – 14 cm-1/%, respectively), for uniaxial tensile [13, 15, 22] and compressive strains [15]. Theory predicts corresponding values to be ∼30 and ∼10 cm-1/% [13, 23]. Moreover, the 2D peak strain sensitivity varies between 45 - 65cm-1/% [13-15, 21, 22], with some samples showing also broadening and splitting of the peak [21, 22, 24]. The observation of 2D splitting depends on the excitation wavelength, the orientation of the graphene lattice along with the direction of strain with respect to incident/scattered light polarization[21, 22, 24-26]. Additionally, there are other parameters which affect the rates of G and 2D wavenumbers as a function of strain of 1LG such as: a) differences in the mechanical set ups, b) use of different substrates, e.g. PMMA, PET, PDMS, c) not efficient stress/ strain transfer and d) slippage of the flake during loading [9, 13-15,18, 21]. Bilayer graphene (2LG) flakes and graphite in uniaxial tension have also been studied by us [16] and others [9] showing similar response to that observed for 1LG.

In this work, the mechanical response of mechanical exfoliated trilayer graphene embedded into polymer substrates under tension and compression is presented by collecting simultaneously G and 2D Raman spectra at each strain level. [16]. In tension the extracted strain rates for both G and 2D peaks are found to be lower in value than previously reported [13-16], although they are similar for both bi- and trilayer parts of the same sample. In compression, the studied flake is an almost isolated trilayer and exhibits a response similar to that reported for the 1LG [27] . Both Raman peaks blue-shift up to a critical compressive strain and the corresponding strain rates within the elastic region of compressive strains are found lower than those expected [15, 27]. The results are discussed and compared with relevant works in the literature.

## 2. Experimental

Graphene bilayer and trilayer flakes were prepared by mechanical exfoliation from natural graphite and transferred onto a PMMA cantilever beam covered by a ~200nm layer of SU8 photoresist. After the deposition and the identification of the flakes using Raman spectroscopy, a thin layer of PMMA photoresist was spin-coated on the top. Two different specimens were investigated, F1 and F2 shown in Fig.1a and b, respectively. The differences in optical contrast on the microphotographs of the flakes indicate thickness variation of different parts on the flake. The brighter areas correspond to thicker parts. Raman spectroscopy revealed that the thicker (brighter) area in flake F1 correspond to a trilayer graphene (3LG) on top of a larger area of bilayer graphene (2LG). Flake F2 consists mainly of a large area 3LG and a thicker (brighter) graphite wedge formed during exfoliation from the transferring tape (fig. 1b).

For the uniaxial mechanical measurements, we used a single cantilever mechanical frame in order to induce the applied tensile and compressive strain to the flakes monitoring the tensile and compressive behaviour of the embedded samples. The top surface of the beam was subjected to tension and compression by flexing down and up the beam, respectively, by means of an adjustable screw positioned at a distance *L* from the fixed end. Raman spectra were collected at discrete values of applied strain.

More information about the cantilever method is presented in references [14, 15].

Raman spectra were measured at 785 nm (1.58eV) using a MicroRaman (InVia Reflex, Rensihaw, UK) set-up. The dashed lines in Fig. 1a,b represent the locations where the Raman mapping took place. The laser power was kept below 1.2 mW on the sample to avoid laser induced local heating. A 100x objective with numerical aperture of 0.85 was used, and the spot size was estimated to be ~1.1 μm [28]. The polarization of the incident light was kept parallel to the strain axis. All Raman spectra were fitted using Lorentzian spectral functions.

## 3. Results and discussion

In sample F1 (Fig.1a) Bernal stacked bilayer and ABA trilayer graphene parts were identified by means of their characteristic Raman features (see also Fig. SI1). In such experiments it is important to determine the transferred strain to the specimen by employing a strain sensor. In this case, we used the well defined strain sensitivity of the Raman G peak of the 2LG flake to derive the actual strain transferred to the whole specimen [16]. To this end, a Raman mapping along a specific line (Fig. 1a) was initially performed on the bilayer part with the sample at rest (0.0%) to define the strain state of the flake prior to the tensile test. The results of the Raman mapping are presented in Fig. 2a where the frequency position (Pos(G)) and full width at half maximum (FWMH(G)) of the G peak as well as the applied strain are plotted as a function of the distance from the right edge of the bilayer (Fig. 1a). Using for zero strain and doped free bilayer graphene the value of Pos(G) ~ 1583.4 cm-1 [29-31], the points along the mapping line were found to redshift to values ranging from 1576.5 to 1583 cm-1. In practice, biaxial strain and unintentional doping is imposed to the flake due to interaction with the substrate and the presence of various adsorbents, and resist/process residuals [29, 32]. Moreover, the biaxial strain distribution on the flake results from an interplay between the forces acting on it, namely, the interaction with the covering PMMA film during sample preparation (spin coating and solvent drying)[14, 15, 30] and the adhesion with the underlying SU8 substrate. The coexistence of both strain and doping can be clearly assessed by the correlation diagram of Pos(2D21) vs Pos(G). It is well known that the 2D Raman peak of the bilayer graphene should be fitted by four Lorentzians namely 2D11, 2D12, 2D21, and 2D22 [16, 33]. As it is clearly shown in Fig. SI1a, the 2D21 component is the most prominent peak and therefore it is that peak which is used in the Pos(2D21) vs Pos(G) correlation diagram. The obtained linear dependence of the Pos(2D21) upon Pos(G) (Fig. 1b) indicates that the interrogated locations are mechanically loaded and doping-free. It should be noted here that for pure mechanically loaded and undoped bilayer graphene the slope in the Pos(2D21) – Pos(G) correlation diagram of Fig. 2b is expected to be 2.2-2.5 [31]. In the examined bilayer graphene the slope of the Pos(2D21) - Pos(G) graph is found to be 2.4, which further corroborated that the mapping points are almost doping free. Thus, as mentioned earlier, the G peak will be used from now on as a biaxial strain sensor (dPos(G)/dε ~ −63 cm-1/% [13]) to quantify the residual strain in graphene at zero strain. As shown in Fig. 2a, there is distribution of residual strains along the mapping line with values ranging from −0.02% (compressive) up to 0.12% (tensile).

In Figure 3a,b the characteristic Raman spectra of the G peak as a function of the applied strain, taken from the bilayer and trilayer part of the flake F1 are shown. Due to the applied uniaxial tensile strain, the G peak redshifts and above a certain strain level splits into the G− and G+ components as expected [13, 14]. The onset of G splitting starts at ~ 0.32% and ~ 0.45% for the bilayer and trilayer, respectively. These values are significantly higher than the 0.15% found in monolayer graphene samples[13]. For the bilayer part, the strain rates of Pos(G−) and Pos(G+) are measured to be −25.9 and −9.7 cm-1/%, respectively, as shown in Fig. 3c. The redshift value of the G−component is somewhat lower than that obtained by others [9] while the corresponding value for G+ is similar to that obtained in [16]. For the trilayer part the strain rates of Pos(G−) and Pos(G+) components are −25.5 and −12.6 cm-1/%,respectively, in agreement (within the experimental error) with the values recorded for the bilayer. This result also indicates that the quality of the flake/ polymer interface is approximately the same throughout the whole flake area and therefore the stress efficiency does not vary throughout. It is important to note here that, to our knowledge, the 3LG uniaxial strain dependence of the G peak for relatively high strains (>0.4%) , is reported for the first time here.

In Fig. 4a,b the evolution of the 2D Raman peak with strain for both the bilayer and trilayer parts of the examined flake is presented. Regarding the trilayer graphene part, it is interesting to note that it exhibits a Bernal ABA stacking configuration due to the fact that the profile of 2D peak at 0% of applied strain is symmetric (Fig. 4b). The 2D Raman peak profile in trilayer graphene is strongly dependent on the excitation wavelength and the atomic registry of the three graphene layers [10, 34]. Therefore, with excitation energy of 1.58 eV (785 nm) the two possible stable configurations, namely ABA and ABC stacks, can be easily distinguished by the amplitude of the two most intense components (Fig. SI2) and the FWHM of the peak fitted by a single Lorentzian [11]. The 2D profile of the ABA stacking is more symmetric due to the equal amplitude of high intensity components and has a narrower FWHM of about 62cm-1 similar to 3LG examined in the present work (Fig. 1) [10].

Commenting on the evolution of the 2D peak with tensile strain for both the bilayer and trilayer parts, it can be inferred that for large strains higher than 0.7%, the profiles of the 2D peaks change significantly and become more symmetric so they can be fitted by a single Lorentzian (Fig. SI3). This is due to the relative shift of the sub- peaks forming the 2D profile, which results from the slight misfit of the atomic registry between the graphene layers in both bilayer and trilayer parts under uniaxial tensile strain. Similar trends were observed in a previous work of ours related to bilayer graphene [16] and in the work of Gong et al [12] related to the loss of Bernal stacking in trilayer graphene. Since the 2D peak frequency results from a double resonance process [33], the misfit of the atomic registry alters the electronic structure of the graphene layers and the phonon frequencies involved in this resonance process. Therefore, the relative intensities of the 2D sub-peaks also change, which may subsequently affect the 2D peak profile as a whole. This effect is more pronounced in the 2D profile of the bilayer part at moderate strain levels e.g. 0.32% and 0.45%, where the change in relative intensities of the $2D_{21}$ and $2D_{22}$ sub-peaks contribute significantly to the changes of the 2D profile relative to that at 0% strain (Fig. 4a).

In Fig. 4c the dependence of the four sub-peaks of the 2D peak profile in bilayer part of the examined flake is shown. The three components, $2D_{12}$, $2D_{21}$ and $2D_{22}$, of lower frequency appear to redshift with the same strain rates of about −47.0 cm-1/%, while for the highest frequency the $2D_{11}$ sub-peak redshifts at a significantly lower rate of −34.0 cm-1/%. For comparative purposes, the measured strain rates of the 2D sub-peaks are not in agreement with the values of −50 cm-1/% and −29 cm-1/% for lower and higher frequency sub-peaks, respectively, reported earlier [16]. Moreover, using a single Lorentzian component to fit the 2D Raman peak the average strain rates of bilayer and trilayer parts can be compared. The strain rates were found to be almost identical ~ −42.0 cm-1/% and ~ −43.0 cm-1/% for bilayer and trilayer parts, respectively (Figure 4d). Therefore, an attempt to fit the 2D peak of the trilayer flake using 6 Lorentzian components, according to literature [33, 35], yields a wavenumber shift per strain of ~ 43 cm-1/% for the four lower frequency components and ~ 37 cm- 1/% for is the remaining two. (Fig. SI4).

Figure 5a,b shows the evolution of G and 2D Raman peaks of trilayer graphene on the flake F2 under uniaxial compression. The maximum applied strain was −0.7%. As expected both peaks are blue-shifted with a strain rate of 14.8 and 43.5 cm-1/% for G and 2D peaks, respectively (Fig. 5c,d). In contrast to the response of trilayer graphene under tension, an onset of a gradual relaxation occurs for both Pos(G) and Pos(2D) at –0.15% strain with slower rates of −9.5 and −25. 6 cm-1/%, respectively. At −0.4% strain a plateau is reached at wavenumber values of 1582 and 2635 cm-1 for Pos(G) and Pos(2D) respectively, which are in effect close to the values at 0.0% strain. Using the Pos(G) as the strain sensor, and taking into account that Pos(G) ~1581.5 cm-1 for unstrained and undoped graphene [10], the strain at the location on the flake where the Raman measurements were conducted (Fig. 1b) is negligible. Furthermore, no G splitting is observed, while the profile of the 2D peak, at each strain level, does not show any significant change compared to the rest (0.0% strain) (Fig. 5). Doping effects cannot be excluded but their influence in Pos(G) and Pos(2D) should be minor since the mean value of <Pos(2D)> defined by fitting a single Lorentzian to the 2D peak, is linearly related to Pos(G) giving a slope of 2.3 (Fig. SI5), which indicates a pure mechanical contribution [31]. The behaviour in compression of a trilayer graphene embedded into a polymer matrix is also presented for the first time here and resembles the behaviour of 1LG examined earlier [14, 15, 27]. As shown recently, the critical strain to failure for the embedded 1LG is not affected by the flake dimensions due to the presence of the polymer constraint and found to be approximately −0.6% [27]. In the 3LG flake examined in the present work the critical buckling strain was found to be -0.15%, which is one fourth of the value found for 1LG [27]. In theory, 3LG exhibits three orders of magnitude greater bending stiffness than single layer graphene and one would expect a higher critical buckling strain. However, as discussed below cohesive failure in compression between the layers not evidently present in 1LG can account for this behaviour. When the polymer matrix is deformed, stress is transferred by interfacial stress transfer to the two outer graphene layers as shown in fig. 6c. Stress transfer to the inner layer can only take place by shear from the two outer layers (fig. 6c). As has been reported previously, the weakest interface is that between the inner graphene layers which in graphite fail at a shear stress of only 30 KPa [12]. For comparison, the shear stress for interfacial stress transfer between the graphene and the polymer is of the order of 1 MPa for model monolayer graphene composites [20, 28]. It is therefore possible that internal shear between the mid and external layers is also present that should lead to premature cohesive failure due to the weakness of the van der Waals forces that bond the layers together. The strain rates for compressive strains of up to 0.15%, were found as -14.8 and -43.5cm-1/% for G and 2D peaks, respectively (fig.5, Table 1). Within this short range of the compressive strains the trilayer flake can be considered as a Hookean elastic plate[36], with strain rates of the two Raman peaks lower than those expected for an effective flake/polymer interface [9, 20] Thus, the unexpected result in the critical buckling strain when the flake thickness increases can be attributed to the relatively bad quality of graphene/polymer interface. Critical transfer length effects are not considered here since the length along the strain axis of the examined flake is ~ 45 μm (Fig. 1b), large enough to ensure effective strain transfer. Similar conclusions for the quality of the flake/polymer interface of the sample F1 has already been drawn by considering the rates of both Pos(G) and Pos(2D) with tensile strain (Figs 3 and 4c,d). Provided that the length of a graphene flake is greater than the critical transfer length for efficient strain transfer [20], the mean strain rates [<dPos(G)/dε>]n or [<dPos(2D)/dε>]n (n = 2, 3 ) can be used to measure the quality of the flake/ polymer interface [9, 20]. It is, thus, concluded that strain rates below -21and -60 cm-1/% for the rates of G and 2D peaks, respectively, indicate poorer quality of the graphene/polymer interface.

In Table I, a summary of the values of the strain rates [dPos(G+, G-)/dε]n and [<dPos(2D)/dε>]n (n = 2, 3 ) found in the present work is presented along with the relevant values found in the literature for bilayer and trilayer graphenes embedded into polymer matrix.

By comparing the values of the strain rates in Table I the question that arises is the following: is the quality of the polymer/flake interface the only reason for the differences among the various works and samples? Starting from the work of Frank et al [16] we observe that the sample studied was formed by a bilayer graphene which was only a part of a large area single layer that was used as the strain sensor (Fig. 6a). Similarly, in the presented work the trilayer graphene of the sample F1 is also part of a larger area bilayer flake (Fig. 6b). It should be pointed out here that although the strain rates in each part (2LG or 3LG) of the sample F1 are underestimated, they are approximately the same for these layers of different thicknesses (Table 1). On the contrary, the flake studied under compression (sample F2) was an isolated bulk trilayer with constant thickness over the whole area (Fig. 6c). The measured strain rate in the elastic region of strains (< 0.15%) is lower than the expected value of ~ 60 cm-1/% (Table 1). In all the above cases, all specimens were long enough (along the strain axis) to avoid critical transfer length effects. Therefore, the architecture of the embedded flake may have a significant contribution in the strain rates of both G and 2D peaks and consequently in the mechanics of the model composite it reinforces. Results from various sample configurations will be the subject of a future work.

## 4. Conclusions

Bernal stacked bilayer and ABA trilayer graphene parts embedded in polymer beams were studied under uniaxial tensile and compressive loadings. Due to the absence of a monolayer graphene part or an individual flake, bilayer part was used as a strain sensor during loading. Raman mapping along a specific line of the bilayer showed that the points onto the flake are almost doping-free and the frequency shifts of both G and 2D peaks are due to the mechanical strain only. In uniaxial tension the G peak of the 3LG splits clearly into the G− and G+ components but at a higher strain levels than those found in single layer graphene. The mean strain rates of G and 2D peaks of both 2LG and 3LG are quite similar, indicating the quality of the flake/ polymer interface is approximately the same throughout the whole flake area. However, the mean strain rates for both G and 2D peaks of the 2LG were found to be lower than those found previously [13-16]. In particular, the 2D peak mean strain rate is lower by about 14% than the value found in ref. [16]. Additionally, the profiles of the 2D peak for both the 2LG and 3LG for large strains higher than 0.7% lose their symmetric appearance indicating a Bernal stacking distortion. The response in uniaxial compression of a trilayer graphene embedded into a polymer matrix was presented for the first time here. Both G and 2D Raman peaks blue-shiftup to a critical compressive strain of – 0.15% strain, where they start relaxing to their zero strain values due to the out-of-plane buckling failure mode. This value of critical strain was found to be one fourth of the value found in single layer graphene despite the enormous bending rigidity that 3LG exhibits as being thicker than 1LG. Moreover, the corresponding strain rates within the elastic region of compressive strains (< 0.15%) were found lower than those expected. Considering that the examined samples were long enough (along the strain axis) to avoid critical transfer length effects, the mean strain rates of G and 2D peaks can be used as a measure of the quality of the flake/ polymer interface. The reduced values found in this work, either in tension or in compression may be attributed mainly to the poor quality of the graphene/polymer interface. Taking into account the architecture of the investigated samples, it is concluded that by tailoring the morphology of a few layer graphene, its adhesion properties with the polymer matrix in which it will act as a reinforcing element may be significantly improved. Further work regarding efficient reinforcing of graphene based nanocomposites will be presented in a future work.

**Acknowledgements**

This research has been co-financed by the European Union (European Social Fund- ESF) and Greek national funds through the Operational Program "Education and Lifelong Learning" of the National Strategic Reference Framework (NSRF) - Research Funding Program: ERC-10 "Deformation, Yield and Failure of Graphene and Graphene-based Nanocomposites". The financial support of the European Research Council through the project ERC AdG 2013, "Tailor Graphene"- Grant agreement no: 321124. Georgia Tsoukleri acknowledges the financial support ofHeracleitus-II PhD grant. Finally, J.P., C. G. and K.P. acknowledge the financial support of the Graphene FET Flagship "Graphene-Based Revolutions in ICT And Beyond"- Grant agreement no: 604391.

# Tables

Table 1: Summary of [dPos(G+, G-)/dε]n and [<dPos(2D)/dε>]n (n = 2, 3 ) values in Tension and Compression

| Reference | Coating | Slope (cm-1/%) | | | Maximum applied strain (%) |
|---|---|---|---|---|---|
| | | G- | G+ | 2D | |
| Bilayer [9] | No | – | – | −38.9 ±2.4 | 0.40 |
| Bilayer [9] | Yes | – | – | −53.9 ±2.9 | 0.40 |
| Bilayer [16] | Yes | −31.3 | −9.9 | ~ −50.0 & ~ −29.0* | 0.80 |
| Trilayer [9] | No | – | – | −32.4 ±0.4 | 0.40 |
| Trilayer [9] | Yes | – | – | −46.6 ±9.0 | 0.40 |
| **In this work** | | | | | |
| Bilayer | Yes | −25.9 | −9.7 | ~ −47.0 & ~ −34.0*  <br> ~ −42.0** | 0.83 |
| Trilayer | Yes | −25.5 | −12.6 | ~ −43.0** | 0.83 |
| Trilayer | Yes | | −14.8 *** | ~ − 43.5 *** | −0.65 |

*Fitting using four Lorentzian components: 2D12, 2D21 and 2D22 show similar slope and 2D11 lower one.

**Fitting using a single Lorentzian.

***Compressive strain up to −0.15%, fitting using a single Lorentzian.

**Figures**

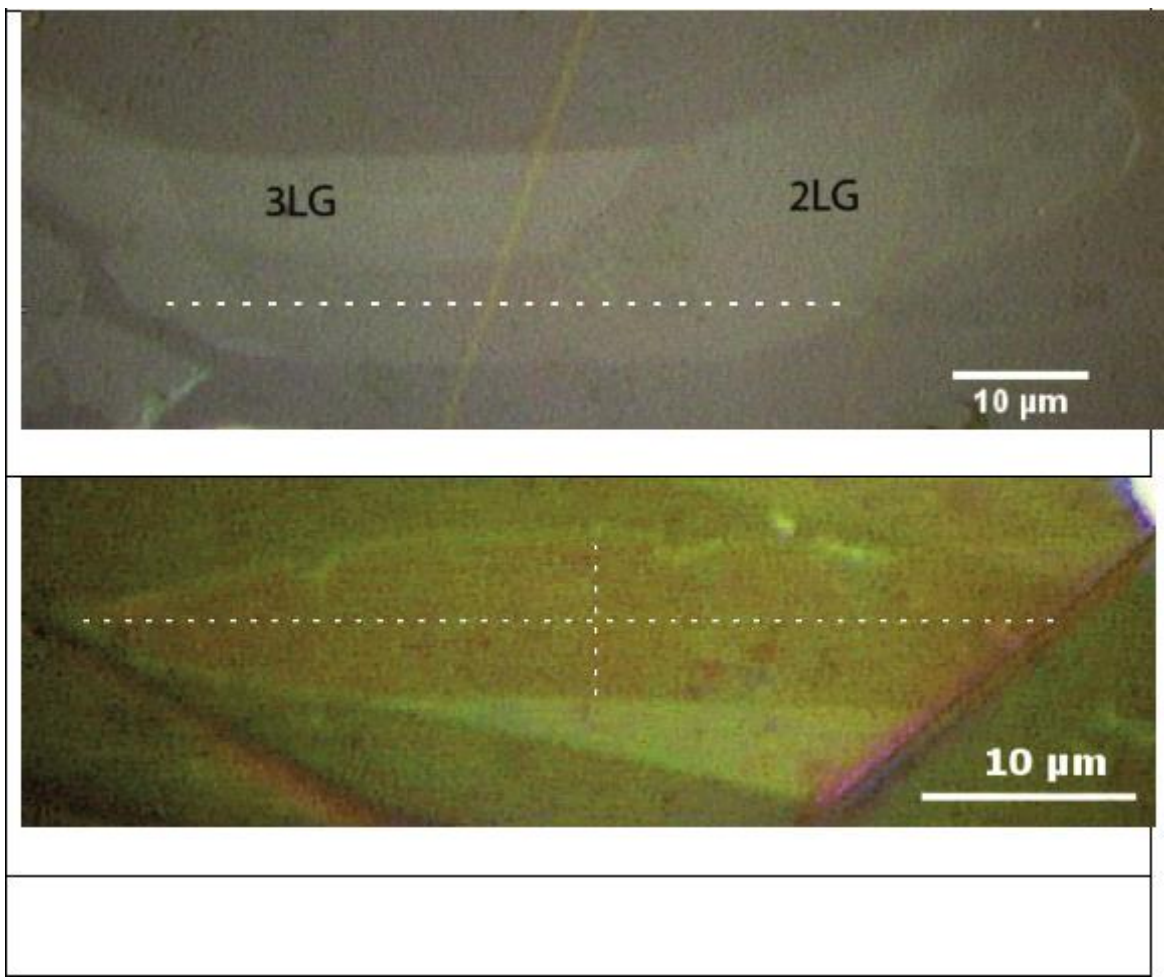

Fig. 1 a) Flake F1 used in the tensile experiment and b) Flake F2 used in the compression experiment. The dashed lines correspond to Raman mapping locations.

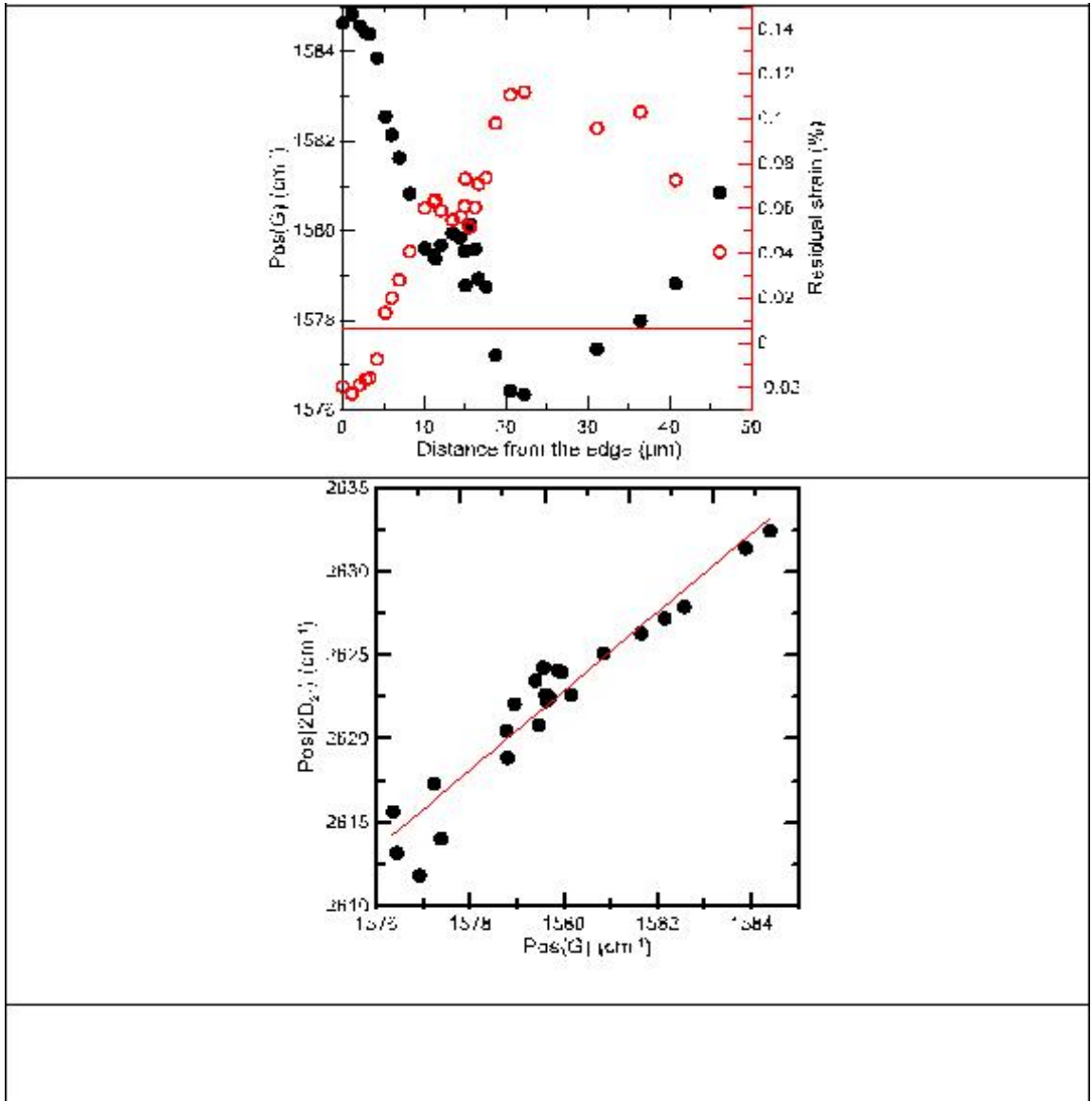

Fig 2 a) Pos(G) and the corresponding residual biaxial strain as a function of the position on the mapping line in Fig.1a, ). Correlation plot of Pos(2D21) vs Pos(G) of the mapping points. The applied strain level is 0%.

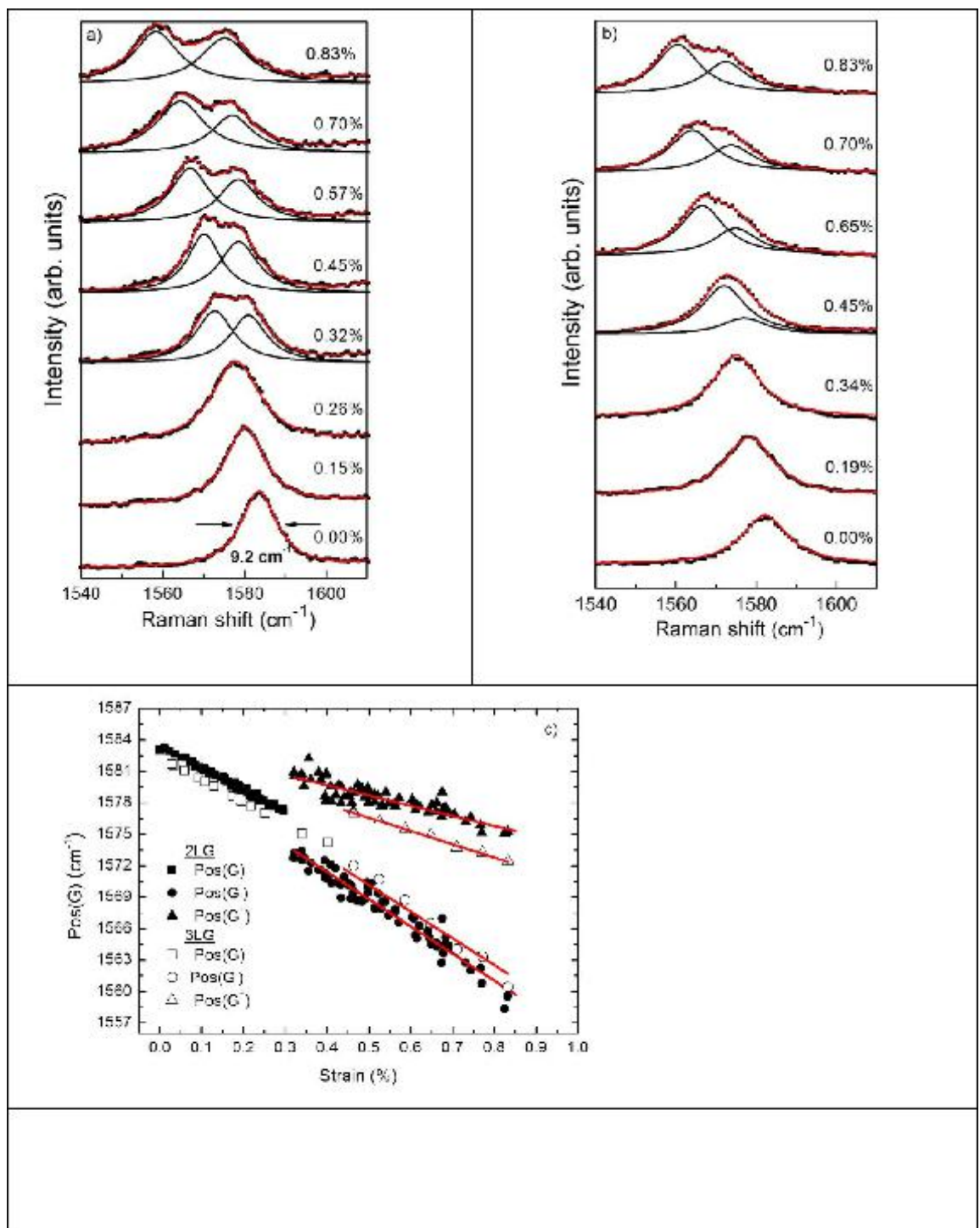

Fig. 3. The evolution of the G Raman peak under uniaxial tensile strain in the (a) bilayer and (b) trilayer part of the F1 flake, (c) Pos(G) as a function of tensile strain for both bilayer and trilayer parts of the examined flake. Raman spectra were acquired from the mapping point shown in Fig. 1.

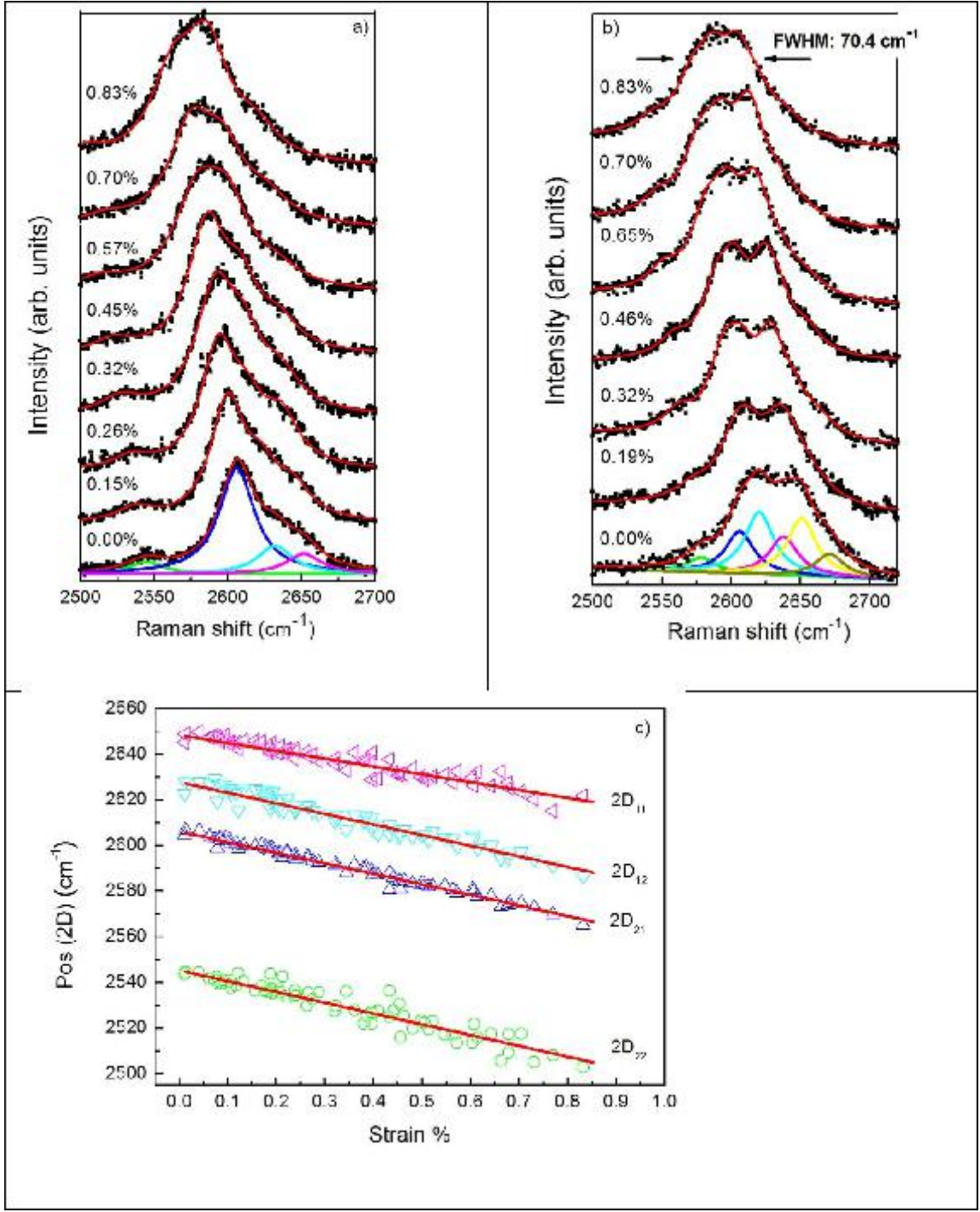

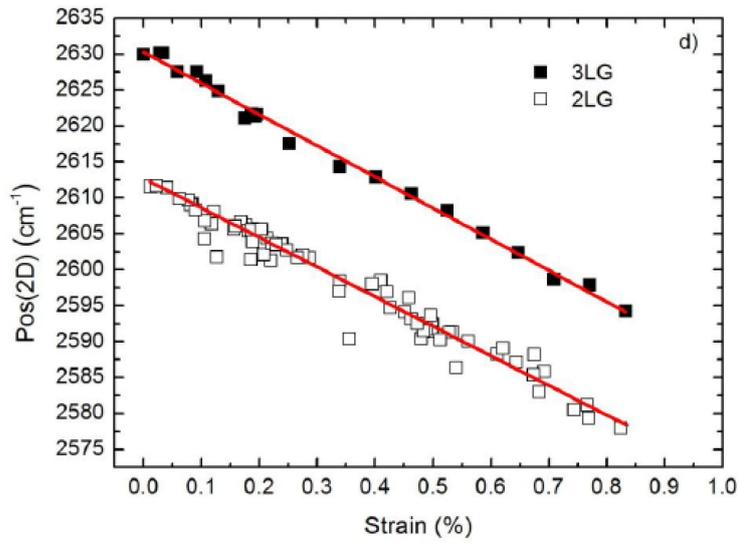

Fig. 4 The evolution of the 2D Raman peak with strain for the (a) bilayer and (b) trilayer part of the examined flake, (c) and (d) Mean Pos(2D) using single Lorentzian fit as a function of strain for bilayer and trilayer part of the examined flake F1.

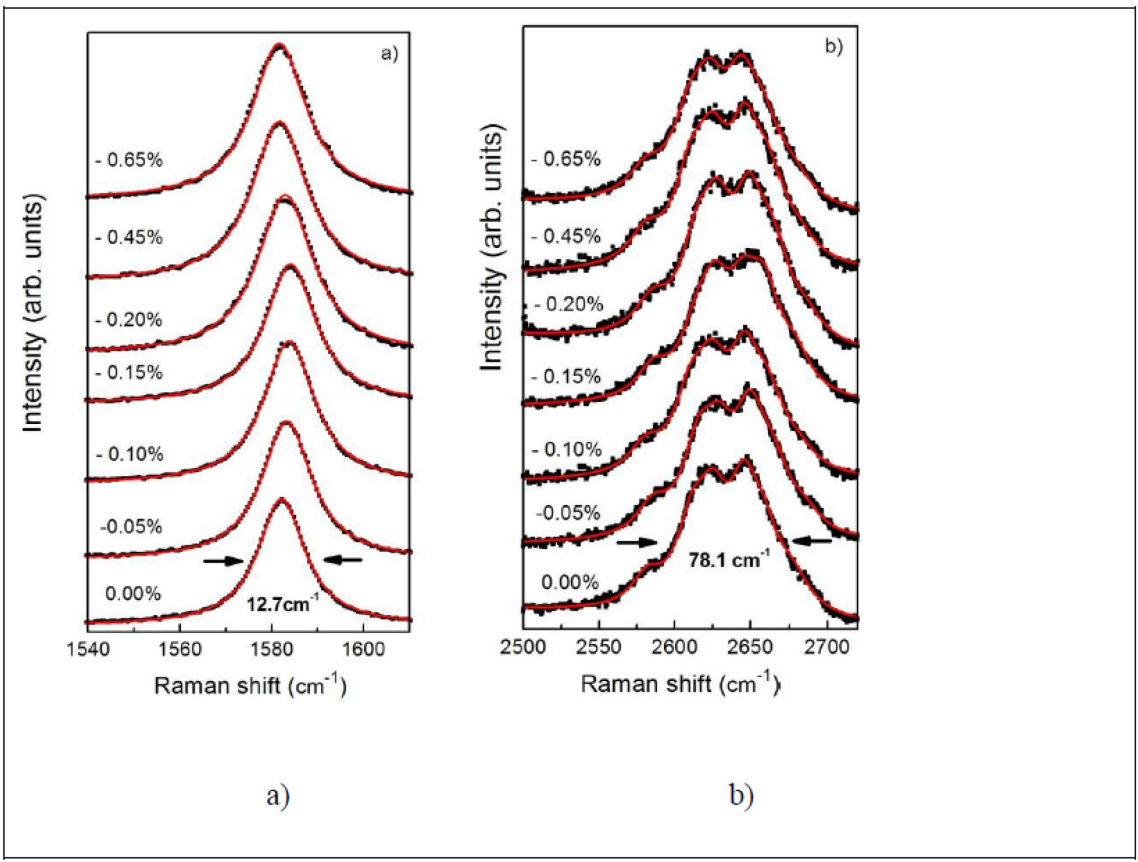

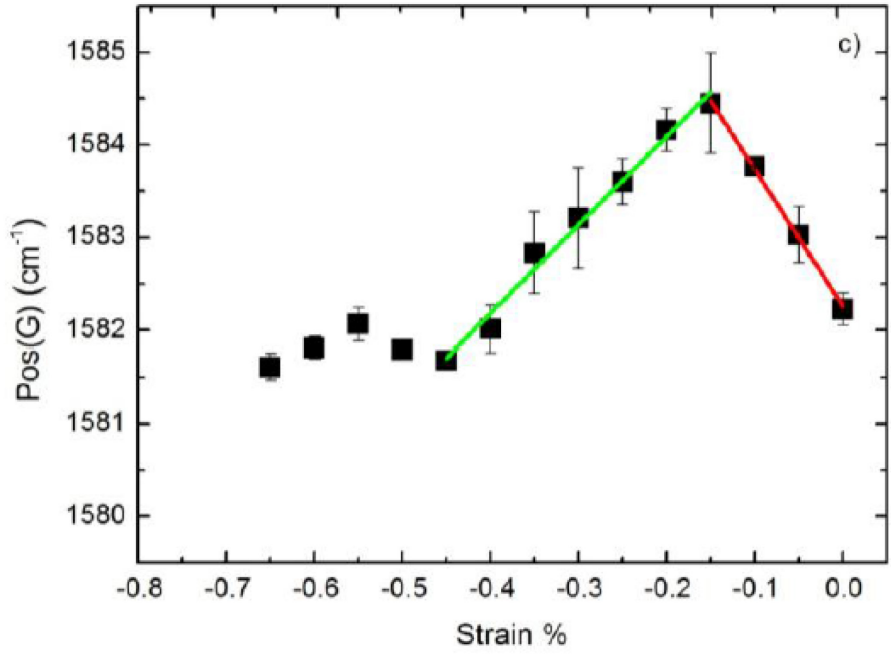

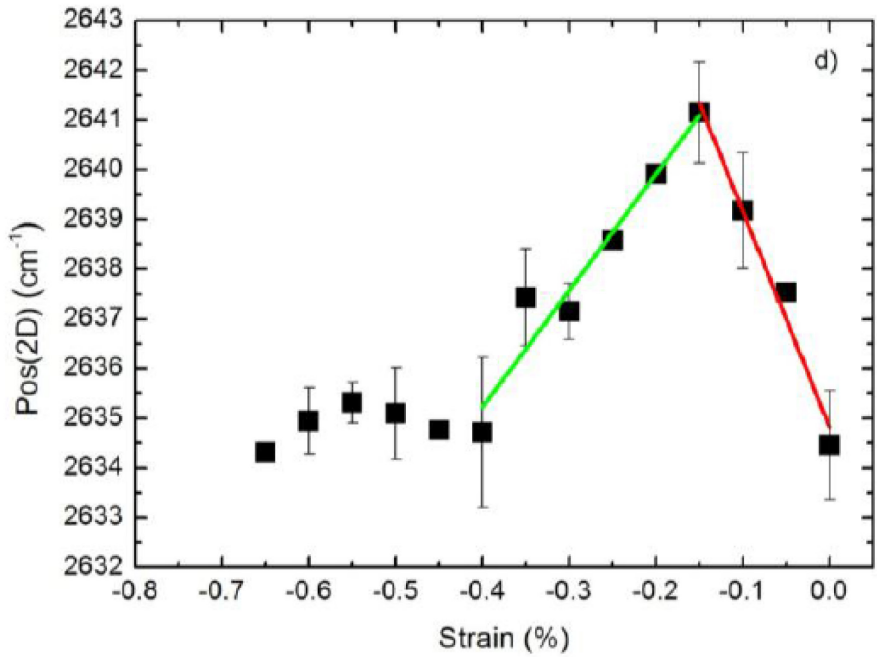

Fig. 5 The evolution of (a) G and (b) 2D Raman peaks and the dependence of (c) Pos(G) and (b) Pos(2D) under uniaxial compression for flake F2.

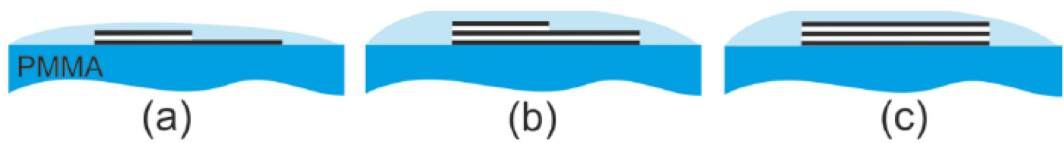

Fig. 6 Schematic representation of various types of few layer graphene samples of different architectures embedded into a polymer matrix. Each black line represents a single graphene layer.